\RequirePackage{lineno}
\documentclass{nature}
\usepackage{xcolor}

\usepackage{caption}
\usepackage{amsmath}
\usepackage{bm}
\usepackage{braket}
\usepackage{amsmath}
\usepackage{amssymb}
\usepackage{textcomp}

\usepackage{bbold}
\usepackage{mathdots}

\title{Population inversion and Dirac fermion cooling in 3D Dirac semimetal Cd$_3$As$_2$}

\author{Changhua Bao$^{1}$, Qian Li$^{1}$, Sheng Xu$^2$,  Shaohua Zhou$^{1}$, Xiang-Yu Zeng$^2$, Haoyuan Zhong$^1$, Qixuan Gao$^1$,   Laipeng Luo$^{1}$, Dong Sun$^{3}$, Tian-Long Xia$^2$ and Shuyun Zhou$^{1,4,\dagger}$}

\usepackage{graphicx}
\makeatletter
\let\saved@includegraphics\includegraphics
\AtBeginDocument{\let\includegraphics\saved@includegraphics}

\makeatother

\begin{document}
\maketitle

\label{abstract}

\begin{affiliations}
	
	\item State Key Laboratory of Low-Dimensional Quantum Physics and Department of Physics, Tsinghua University, Beijing 100084, P. R. China
	\item Department of Physics and Beijing Key Laboratory of Opto-electronic Functional Materials and Micro-nano Devices, Renmin University of China, Beijing 100872, P. R. China
	\item International Center for Quantum Materials, School of Physics, Peking University, 100871 Beijing, P. R. China
	\item Frontier Science Center for Quantum Information, Beijing 100084, P. R. China\\
	
	$\dagger$e-mail: syzhou@mail.tsinghua.edu.cn
\end{affiliations}

\begin{abstract}
Revealing the ultrafast dynamics of three-dimensional (3D) Dirac fermions upon photoexcitation is critical for both fundamental science and device applications.   So far, how the cooling of 3D Dirac fermions differs from that of two-dimensional (2D) Dirac fermions and whether there is population inversion are fundamental questions that remain to be answered.  Here we reveal the ultrafast dynamics of Dirac fermions in a model 3D Dirac semimetal Cd$_3$As$_2$ by ultrafast time- and angle-resolved photoemission spectroscopy (TrARPES) with a tunable probe photon energy from 5.3 - 6.9 eV. The energy- and momentum-resolved relaxation rate shows a linear dependence on the energy, suggesting Dirac fermion cooling through intraband relaxation.  Moreover, a population inversion is reported based on the observation of accumulated photoexcited carriers in the conduction band with a lifetime of $\tau_n$ = 3.0 ps. Our work provides direct experimental evidence for a long-lived population inversion in a 3D Dirac semimetal, which is in contrast to 2D graphene where the interband relaxation occurs on a much faster timescale.

\end{abstract}

\renewcommand{\thefigure}{\textbf{Fig. \arabic{figure} $\bm{|}$}}
\setcounter{figure}{0}


\label{intro}

Three-dimensional (3D) Dirac semimetals  \cite{ArmitageRMP2018,DingHRMP2021} are characterized by Dirac fermions with linear dispersion near the point-like Dirac nodes in the 3D momentum space, which can be viewed as 3D analogues of two-dimensional (2D) Dirac fermions in graphene \cite{geim2007,NetoRMP09}. Such unique electronic structure  strongly affects the light-matter interaction, and can lead to intriguing optoelectronic properties with potential applications  in devices such as high-performance photodetectors \cite{SunDRev}.
Moreover, the vanishing density of states (DOS) near the Dirac point suppresses the scattering phase space, and the slower interband relaxation could lead to population inversion, which is highly desirable for broad-bandwidth Terahertz or mid-infrared lasing application \cite{weber2021} and realizing exotic dynamical excitonic insulator \cite{Alexander2020}. So far, Dirac fermion cooling dynamics has been widely investigated for 2D graphene, and a population inversion with a timescale of $\sim$130 fs has been reported \cite{Cavalleri2013}.  However, how the cooling of 3D Dirac fermions differs from 2D Dirac fermions and whether it is possible to host a long-lived population inversion are fundamental questions that are still awaiting to be answered.

As a model 3D Dirac semimetal, Cd$_3$As$_2$ has attracted extensive research interests  \cite{FZ2013,Chen2014,Hasan2014,Cava2014,Ong2015} with a large Fermi velocity, high carrier mobility, good chemical stability, and it shows broadband responses for ultrafast nano-optoelectronics \cite{Sun2017_2,Zhu2017}. The understanding of the fundamental ultrafast carrier dynamics upon photoexcitation is therefore important for its nano-optoelectronics device applications.
 So far, the ultrafast dynamics of 3D Dirac fermions have been investigated by time-resolved optical measurements \cite{Alex2015,Sun2017,Sun2017_2,Wang2017,Zhao2017,Zhu2017,Sun2018,Ma2019,Zhang2020}, and two main relaxation processes involving a faster one in 0.4 to 4 ps \cite{Alex2015,Sun2017} and a slower one from 2 to 8 ps \cite{Sun2017,Alex2015,Sun2018} have been reported. Recently, a possible long-lived population inversion has been suggested from the ultrafast response of THz photoconductivity \cite{Ma2019}. We note that the optical response is contributed by all photocarriers at different electron energy and momentum, making it difficult to identify the contribution of electrons from different parts of the Dirac cone. Resolving the Dirac fermion dynamics with both energy- and momentum-resolved information will allow to extract the chemical potential, electronic temperature, energy-resolved relaxation rate etc., which are critical for understanding the Dirac fermion cooling dynamics and confirming whether there is population inversion.  In addition, photo-induced topological phase transitions have been proposed for 3D topological semimetals \cite{Oka2016,Lee2016,Rubio2017,weber2021,NRP2021}, and accessing the dynamics of such 3D topological semimetals would be a first important step.

Time- and angle-resolved photoemission spectroscopy (TrARPES) is a powerful technique for revealing the ultrafast dynamics of Dirac fermions with both energy- and momentum-resolved  information \cite{Lanzara2016,ShenRMP2021}. While this technique has been applied to various quasi-2D materials such as high-temperature superconductors \cite{Wolf2007,Lanzara2011,Bovensiepen2012}, topological insulators \cite{ZX2012,Gedik2012} and graphene \cite{Cavalleri2013,Hofmann2013} to reveal rich dynamics information,  TrARPES study of 3D Dirac fermions has been missing so far due to the lack of widely tunable probe photon energy, which is required to scan through the conical dispersion at the desired $k_z$ precisely.
Here by using TrARPES with a tunable probe photon source from 5.3 to 7.0 eV, we are able to reveal the transient dynamics of 3D Dirac fermions in Cd$_3$As$_2$ upon photoexcitation.
Our TrARPES measurements reveal two dominant processes in the 3D Dirac fermion cooling, namely intraband relaxation through electron-phonon scattering as indicated by a linear energy-dependent relaxation rate, and interband relaxation indicated by accumulated Dirac fermions in the conduction band (CB). Such accumulated Dirac fermions in the CB suggests a population inversion with a lifetime of $\tau_n$ = 3.0 ps.  Our work provides direct experimental evidence for a long-lived population inversion in Cd$_3$As$_2$, and opens up opportunities for exploring the light-matter interaction for 3D Dirac fermions.

\begin{figure*}[htbp]
	\centering
	\includegraphics[width=15.5 cm]{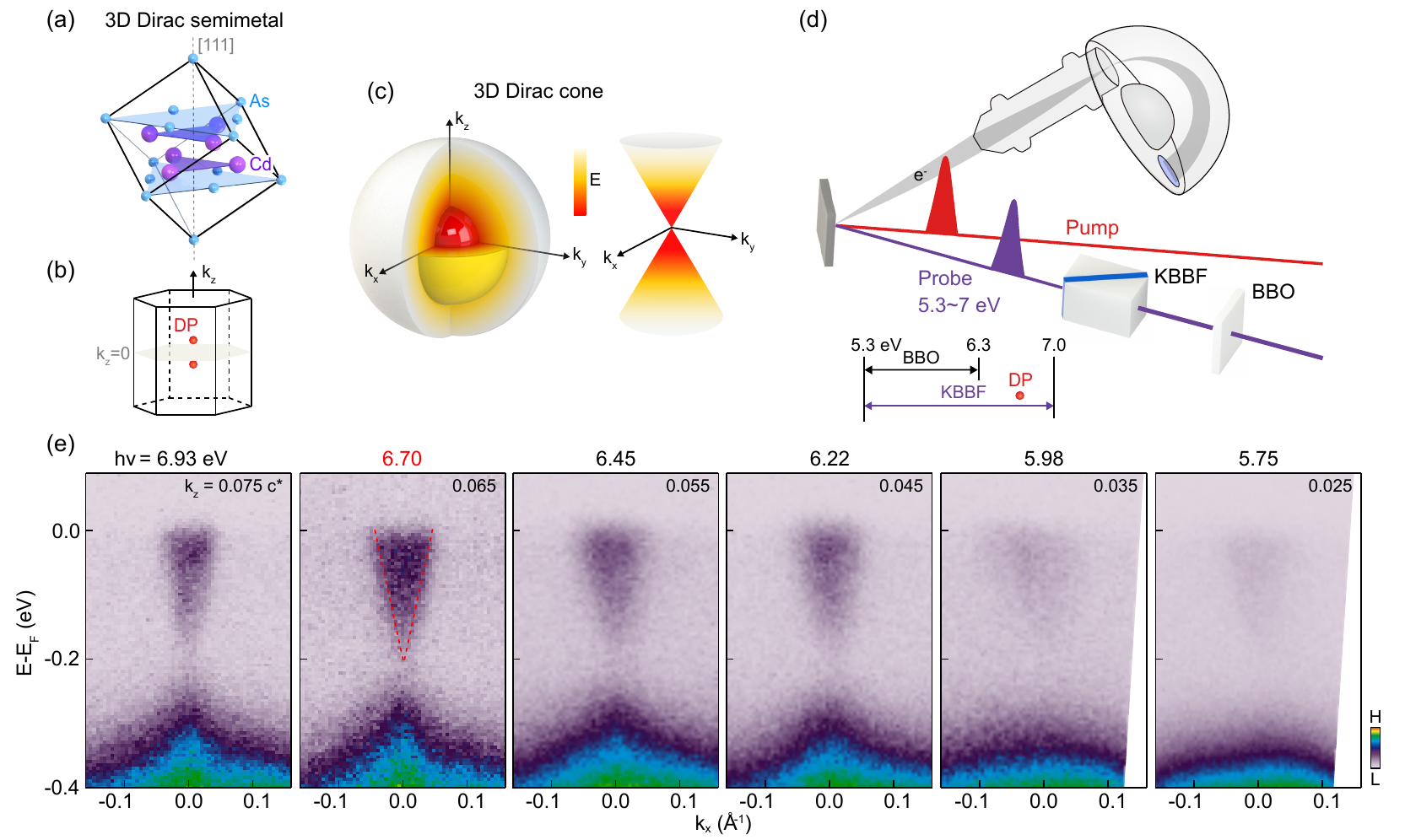}
	\caption{\textbf{Revealing the 3D Dirac cone by TrARPES with tunable probe photon energy.} (a) Crystal structure of Cd$_3$As$_2$. (b) Hexagonal Brillouin zone of Cd$_3$As$_2$ and Dirac nodes. (c) Schematics of 3D Dirac cone and its projection in the $k_x$-$k_y$ plane. (d) A schematic drawing of time-resolved ARPES with tunable probe photon energy. The inset shows the photon energy range by using BBO and KBBF based fourth harmonic generation and the corresponding photon energy for 3D Dirac point. (e) Dispersions along the $\Gamma$-M direction with photon energy from 6.93 to 5.75 eV. An inner potential V$_0$ = 10.6 eV is used to calculate $k_z$.}
\end{figure*}

Cd$_3$As$_2$ crystallizes in the cubic structure as shown in Fig.~1(a).  3D Dirac cones emerge at two isolated Dirac nodes at $k_z$ = $\pm$0.065 c$^*$ in the $k_x$-$k_y$-$k_z$ momentum space \cite{Hasan2014,Cava2014,Chen2014} as indicated by dots in the hexagonal Brillouin zone (BZ) in Fig.~1(b). The electronic structures near these Dirac nodes are schematically illustrated in Fig.~1(c), where a linear dispersion is expected  when cutting through the Dirac node.  In order to access the Dirac node, a tunable probe photon energy covering the right $k_z$ value is required.  A calculation using an inner potential of 10.6 eV \cite{Chen2014} shows that $k_z$ of the Dirac node (0.065 c$^*$) corresponds to a probe photon energy of 6.7 eV. A TrARPES system with a tunable probe photon energy covering 6.7 eV is therefore required for revealing the dynamics of the Dirac electrons in Cd$_3$As$_2$.

In TrARPES, the maximum photon energy generated from fourth harmonic generation (FHG) of BBO crystal is 6.3 eV \cite{Wolf2007,Silvestri2009}, which is not sufficient to access the Dirac node yet.  KBBF crystal can extend the FHG output to 7.0 eV \cite{Chen2008}, and a continuously tunable photon source covering 5.6 - 7.0 eV has been used in static laser-based ARPES measurements \cite{Kaminski2014,ZhouXJRPP2018}.  Here we develop its time-resolved version to facilitate ultrafast dynamic studies of 3D materials as shown in Fig.~1(d) (see more details in ref~ \cite{ZhouRSI2021}).
A laser beam with a tunable wavelength of 710 - 930 nm and a pulse duration of 50 - 80 fs is generated from a Ti:sapphire oscillator and split into pump and probe beams.  The latter is further converted into a tunable photon energy of 5.3 - 7.0 eV (wavelength of 232.5 - 177.5 nm) by KBBF-based FHG, allowing to probe the carrier dynamics near the 3D Dirac node.

Figure 1(e) shows ARPES dispersion images measured at photon energies from 6.93 to 5.75 eV on naturally-cleaved (111) surface of Cd$_3$As$_2$  at 80 K (see the Supplemental Materials). The conical-shaped dispersion is clearly observed at 6.70 eV, whose intensity decreases when detuned from this photon energy and becomes undetectable at 5.75 eV, confirming its 3D nature.  The Dirac cone appears at the photon energy of 6.70 eV  with the corresponding $k_z$ = 0.065 c$^*$, consistent with previous ARPES studies using synchrotron light sources \cite{Hasan2014,Cava2014,Chen2014}.

\begin{figure*}[htbp]
	\centering
	\includegraphics[width=15.5 cm]{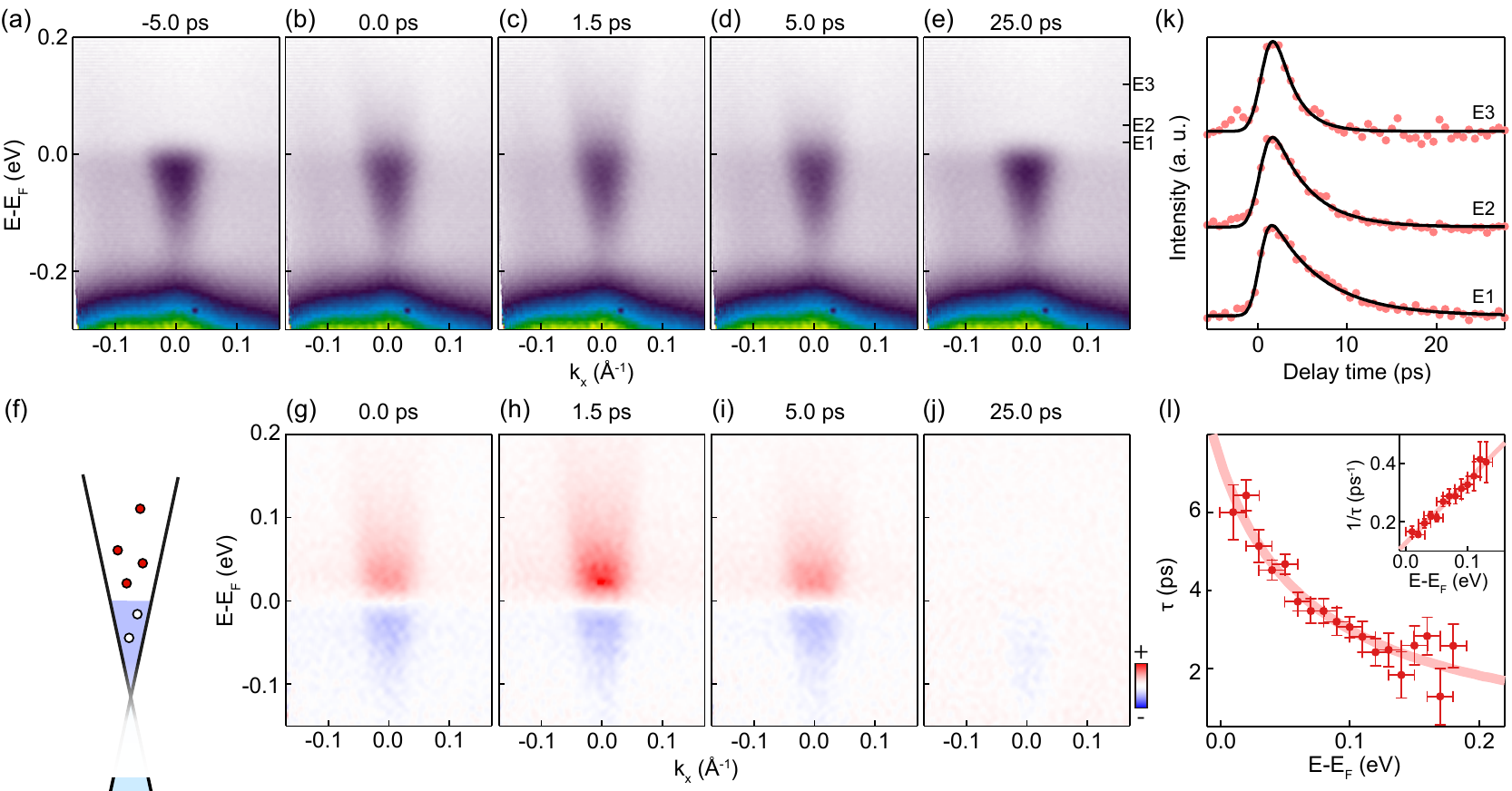}
	\caption{\textbf{Photoexcited carrier dynamics of 3D Dirac fermions.} (a)-(e) TrARPES spectra at different delay times. (f) Schematics for photoexcited 3D Dirac fermions. (g)-(j) Differential TrARPES spectra at different delay times after subtracting the spectrum at -5 ps.  (k) Temporal evolution of photoexcited Dirac fermions at a few selected energies indicated by tick marks in (e). (l) Extracted lifetimes $\tau$ of photoexcited Dirac fermions as a function of  energy. The inset shows the scattering rate $1/\tau$ as a function of energy with a linear fitting. The pump and probe photon energies are 6.7 and 1.68 eV respectively, with a pump fluence of 40 uJ/cm$^2$.}
\end{figure*}

Using the selected probe photon energy of 6.70 eV, the dynamics of 3D Dirac fermions is  revealed by TrARPES measurements. Figures 2(a)-(e) show TrARPES snapshots of the Dirac cone at different delay times. Photoexcitation of electrons into the unoccupied states up to 0.2 eV above $E_F$ [indicated by red color in the differential images in Fig.~2(g)-(j)] and photoexcited holes below $E_F$ [indicated by blue color in Fig.~2(g)-(j)] are observed as schematically illustrated in Fig.~2(f).  The photoexcited electrons above 0.1 eV relax much faster and recover at 5 ps [Fig.~2(d)], while those near $E_F$ return to the equilibrium state at a later delay time and completely recovers by 25 ps [Fig.~2(e)].
The lifetimes of photoexcited carriers at different energies can be extracted by fitting the temporal evolution of TrARPES spectra as shown in Fig.~2(k). A single-exponential function is sufficient to fit the data. The extracted energy-dependent lifetime of photocarriers excited above $E_F$ is plotted in Fig.~2(l), which increases from 2 ps near 0.2 eV to 6 ps near $E_F$. Considering that the coupling between electron and acoustic phonon typically leads to a much longer lifetime in Cd$_3$As$_2$ \cite{Fiete2015} than what is observed here, the main path for Dirac fermion cooling is through coupling to optical phonons. The electron-phonon (el-ph) coupling has been shown to play an important role for the carrier dynamics in topological materials \cite{Burch2021PRX,Jaime2021CP}. It is worth noting that such carrier lifetime is much longer than the lifetime reported in 2D Dirac fermion in graphene (0.25 ps) \cite{Elsaesser2011} because of the lower optical phonon energy in Cd$_3$As$_2$ \cite{Lemmens2017}.

To extract the el-ph coupling strength $g$, we replot the scattering rate (inversely proportional to the relaxation lifetime) as a function of energy in the inset of Fig.~2(l). A linear energy dependence of scattering rate is clearly identified, and the electron-phonon coupling strength $g$ can be extracted from the slope, which equals to $\frac{g(T_e-T_L)}{2C_e k_B T_e}$ \cite{Wu2012,Wang2017}, where $C_e$ is the electron heat capacity, $T_e$ is the transient electronic temperature and $T_L$ is the lattice temperature. The slope of $2.1\pm0.1$ ps$^{-1}\cdot$eV$^{-1}$ gives an el-ph coupling strength $g$ = $(6.2\pm 0.3)\times10^{15}$ $W\cdot K^{-1}m^{-3}$, where $C_e$ = 70 $J\cdot K^{-2}m^{-3} \cdot T_e$ \cite{Wang2017}, $T_e$ = $450$ K (extracted from the Fermi-Dirac distribution) and $T_L$ = $80$ K (sample  temperature). We note that similar wavelength-dependent relaxation rates with a more complex relation deviating from linearity have been observed by time-resolved optical measurements \cite{Zhu2017,Wang2017,Zhang2020} where different electronic states in the Dirac cone are selectively excited by tuning the photon energy.  Here, the direct energy- and momentum-resolving capability allows to access each electronic state in the 3D Dirac cone directly and to extract the el-ph coupling strength more accurately.

\begin{figure*}[htbp]
	\centering
	\includegraphics[width=15.5 cm]{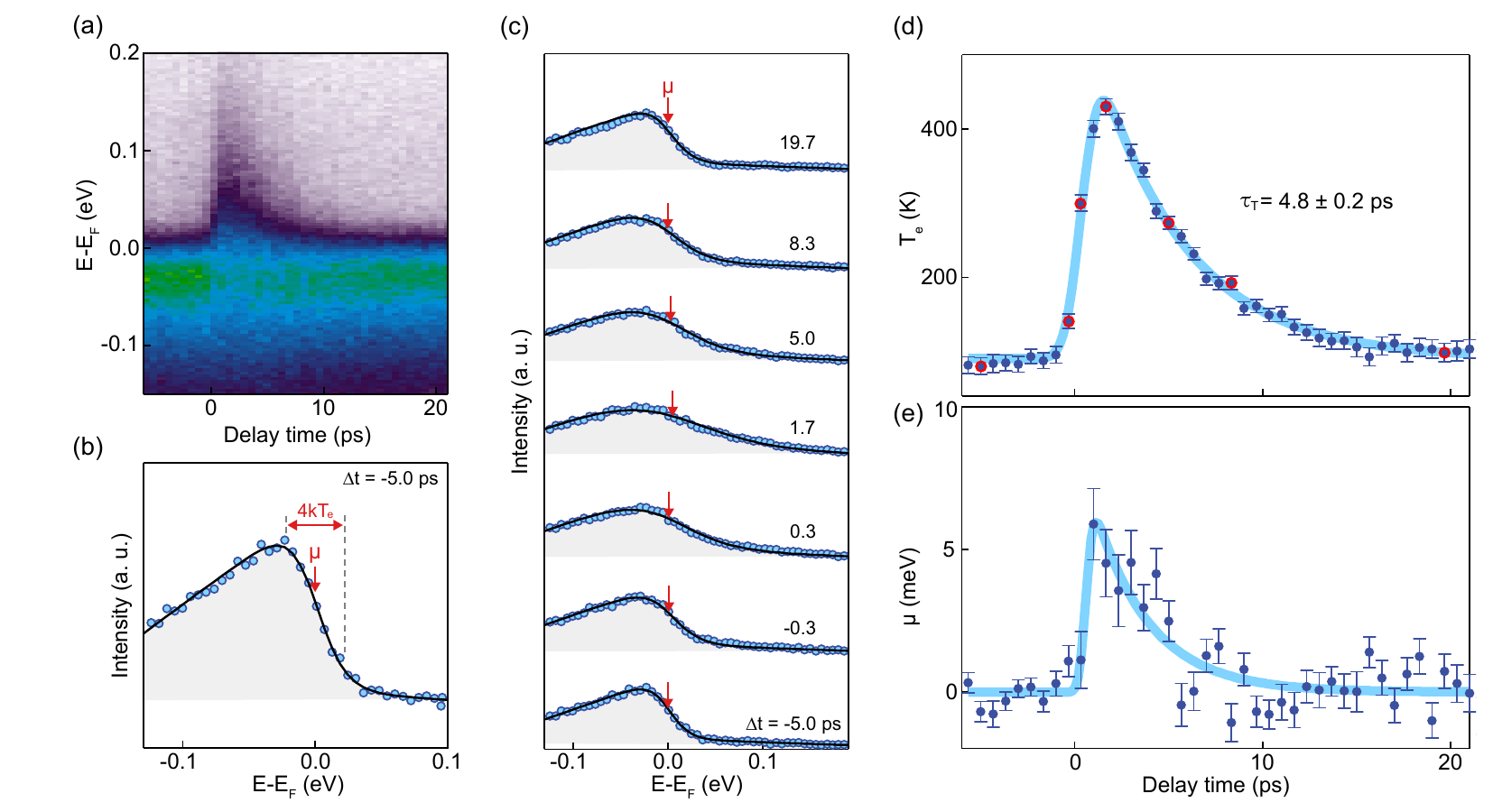}
	\caption{\textbf{Ultrafast thermodynamic evolution of 3D Dirac fermions.} (a)  Integrated EDCs as a function delay time. (b)  EDC at delay time of -5.0 ps and corresponding Fermi-Dirac fitting. (c) Representative EDCs at delay times of -5.0, -0.3, 0.3, 1.7, 5.0, 8.3 and 19.7 ps and black curves are Fermi-Dirac distribution fittings. (d) Extracted electronic temperature as a function of delay time. The data points extracted from the curves in panel (c) are highlighted by red color. (e) Extracted chemical potential as a function of delay time.}
\end{figure*}

Such direct snapshots of photoexcited 3D Dirac fermions allow to reveal not only the photo-carrier dynamics but also the thermodynamic evolution \cite{Hofmann2014,Bauer2018,Murnane2019}, which could provide further insight into the underlying physics, in particular whether there is population inversion.  Figure 3(a) shows the evolution of photoelectrons as a function of energy and delay time. Thermodynamic parameters including the transient electronic temperature $T_e$ and chemical potential $\mu$ can be extracted by fitting the integrated energy distribution curves (EDCs) [Fig.~3(a)] with a Fermi-Dirac distribution as illustrated in Fig.~3(b), where the edge position defines $\mu$ while the width of the edge is determined by 4$kT_e$. Figure 3(c) shows EDCs at a few selected delay times, and the extracted $T_e$ and $\mu$ are plotted in Fig.~3(d),(e).
The cooling of the electronic temperature is through the energy transfer from hot electrons to the cold lattice, which could be described by the two-temperature model (TTM) \cite{allen1987} $C_e\frac {dT_e}{dt}=-g(T_e-T_L)$, the integration of which gives $T_e(t)=(T_e(0)-T_L) e^{-\frac {t} {C_e/g}}+T_L$ where $T_e(0)$ is electronic temperature at $t=0$. Analysis of the electronic temperature shows a single-exponential relaxation consistent with TTM and gives a relaxation time $\tau_T$ of 4.8$\pm$0.3 ps, providing the important characteristic time of el-ph scattering in Cd$_3$As$_2$. In addition, the el-ph coupling strength $g$ can also be extracted from the relaxation time $\tau_T$ by $g$ = $C_e/\tau_T$ = $(6.6\pm 0.4)\times10^{15}$ $W\cdot K^{-1}m^{-3}$, which is consistent with the value extracted from photo-carrier dynamics discussed above, thus confirming the validity of the thermodynamic analysis.

\begin{figure*}[htbp]
	\centering
	\includegraphics[width=15.5 cm]{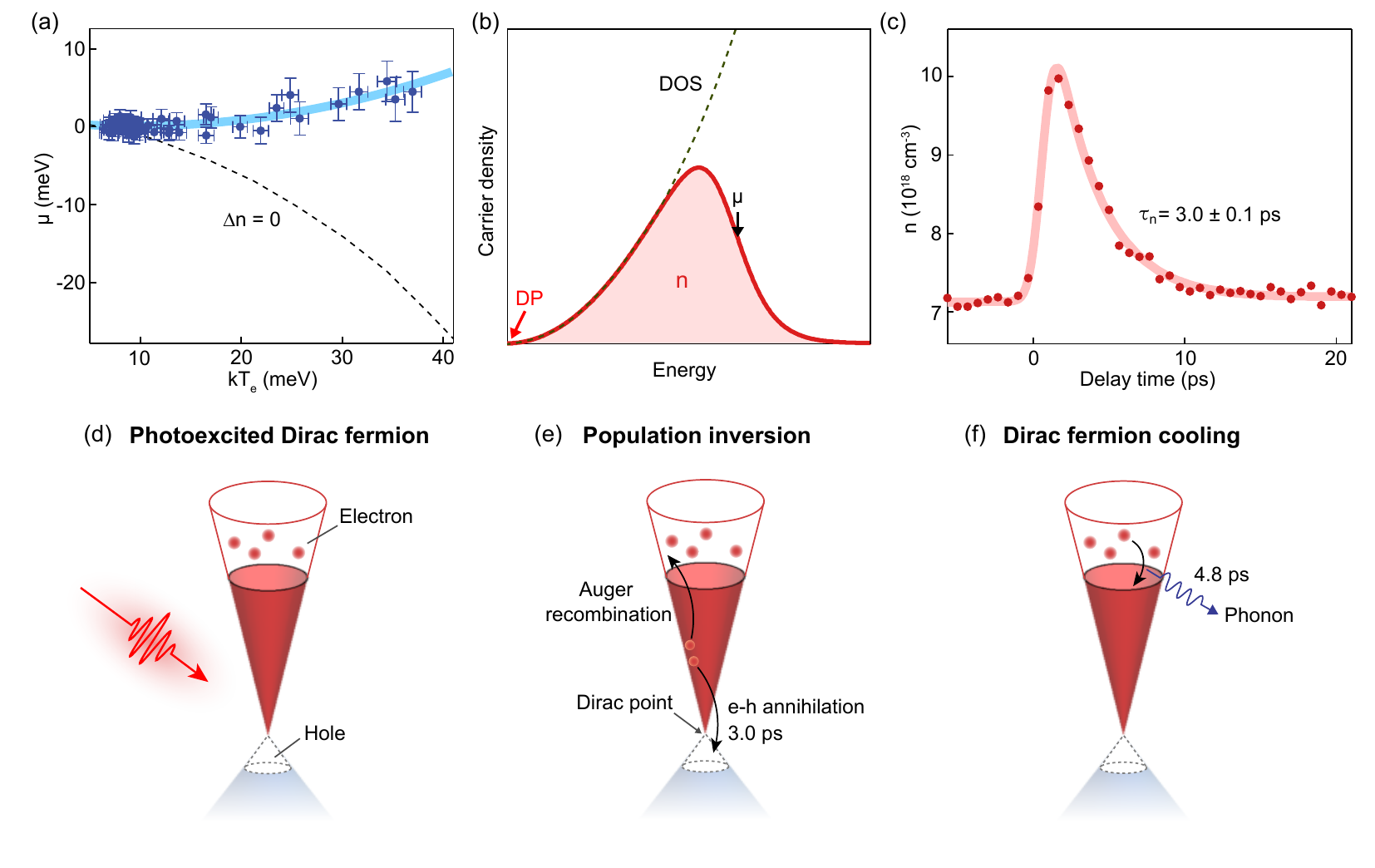}
	\caption{\textbf{Experimental evidence of population inversion and schematic 3D Dirac fermion relaxation mechanism.} (a) Replotted $\mu$ as a function of $kT_e$. The black dashed curve is simulated temperature dependent chemical potential for 3D Dirac fermion under the assumption of carrier conservation. (b) Simulated energy dependent carrier density of 3D Dirac semimetal to illustrate the calculation of carrier concentration in CB. (c) Carrier concentration in CB as a function of delay time.  (d)-(f) Schematics for 3D Dirac fermion photoexcitation, population inversion and cooling mechanism.}
\end{figure*}

The chemical potential also shows a shift with delay times, which  can be used to check if there is any population inversion when combined with the electronic temperature and density of states analysis.  The increase of the chemical potential is identified from the shift of the edge in Fig.~3(c) and the extracted values with a maximum value of 6 meV are plotted in Fig.~3(e). To check if the increased chemical potential implies the population inversion, three possible mechanisms are examined, including electron-hole asymmetry, heating effects and carrier concentration change.  First, electron-hole asymmetry \cite{Alessandra2015PRB} can be excluded, because the thermal excitation energy ($kT_e=39~meV$ for $T_{e, max} = 450~K$ as shown in Fig.~3(d)) is much smaller than the binding energy of the Dirac point ($\sim$200 meV), and therefore no electron can be excited thermally from the VB to CB and no hole can be generated in VB.  Secondly, the heating effects can also be excluded. The increase of T$_e$ means that more electrons are excited to higher energy, meanwhile the DOS for the CB increases with the energy for 3D Dirac fermions as $\rm{DOS}(E) \propto E^2$.  If the number of carriers is conserved for the CB, a lower $\mu$ would be expected (see simulation in Fig.~4(a)), which is in contrast to our experimental results. Therefore, the increase of chemical potential suggests the carrier concentration for the CB of the 3D Dirac cone is not conserved upon photoexcitation, namely, there is a population inversion with photoexcited electrons in the CB.

To confirm the population inversion, we further extract the temporal evolution of the carrier concentration.  The carrier density at each energy is determined by $T_e$ and $\mu$ through $\rho(E) = \frac {(E-E_{DP})^2}{\pi^2\hbar^3 v_x v_y v_z}\frac{1}{1+e^{(E-\mu)/kT_e}}$, which is schematically shown in Fig.~4(b).  The total carrier concentration at each delay time can be obtained by integrating $\rho(E)$ over energy $n=\int_{E_{DP}}^{\infty} \rho(E){dE}$ using the corresponding extracted $T_e$ and $\mu$ in Fig.~3(d),(e) (see the Supplemental Materials). The temporal evolution of carrier concentration in Fig.~4(c) shows a clear increase of carrier concentration in the CB.  We note that the change of carrier concentration in the VB, however, cannot be direct extracted from the TrARPES data due to the multi-band nature of the VB \cite{Crepaldi2018} and the complications from the other strong band around -0.4 eV, which exists in different $k_z$ values and dominates the intensity of the VB over the intrinsic 3D Dirac cone. Therefore, the change in the carrier concentration of the VB has to be deduced indirectly by considering the change in the VB together with the conservation of total charge carriers.  The conservation of total charge carriers is based on the fact that the pump photon energy (1.68 eV) is much smaller than the work function (typically around 4.5 eV), and therefore no electrons can be excited out of the crystal. Therefore, the observation of increased carrier concentration in CB is equivalent to the increase of the same number of holes in VB due to the conservation of carriers, thus confirming the light-induced population inversion. Fitting the relaxation curve in Fig.~4(c) shows that the population inversion in Cd$_3$As$_2$ has a lifetime of 3.0$\pm$0.1 ps, which is an order of magnitude longer than the population inversion lifetime (130 fs) in graphene \cite{Cavalleri2013}. This suggests that the interband relaxation plays an important role in 3D Dirac fermion cooling in addition to el-ph scattering induced  intraband relaxation upon photoexcitation as summarized in Fig.~4(d)-(f).  However, the interband relaxation for 3D Dirac fermions is much slower compared to 2D Dirac fermions \cite{Gedik2012,Cavalleri2013}, where the population inversion exists only in 130 fs \cite{Cavalleri2013}.

The long-lived population inversion in 3D Dirac semimetal is likely related to the suppression of Auger recombination, namely, electron-hole recombination with energy and momentum transfer to another carrier (schematically illustrated in Fig.~4(d)).
The Auger recombination probability is determined by both the scattering phase space and the scattering matrix element.  For 2D Dirac fermions, although the scattering phase space is vanishing, the scattering matrix element is diverging \cite{Ermin2010}.  This leads to a finite Auger recombination probability, which is a main obstacle for population inversion.  For 3D Dirac semimetals, theoretical work suggests that Auger recombination is suppressed because of the vanishing phase space and the finite scattering matrix element \cite{Svintsov2019}.  This could lead to order of magnitude enhancement in the lifetime of population inversion as reported in our TrARPES data.

In conclusion, by utilizing the TrARPES system with widely tunable probe photon energy to reveal the photoexcited 3D Dirac fermions directly, the relaxation dynamics of the 3D Dirac fermions and its distinction from 2D Dirac fermions is revealed.  Moreover, a population inversion with a lifetime of 3.0 ps is reported.  Our work reveals the fundamental dynamics of 3D Dirac fermions, and opens up new opportunities for exploring the dynamics of topological semimetals as well as light-induced transient topological phase transitions.


\begin{addendum}
	\item[Acknowledgements] This work is supported by the National Key R$\&$D Program of China (Grant Nos.~2020YFA0308800 and 2021YFA1400100), the National Natural Science Foundation of China (Grant Nos. 11725418 and 11427903). T.-L.X. is supported by the National Natural Science Foundation of China (Grant Nos.~12074425 and 11874422) and the National Key R$\&$D Program of China (Grant No.~2019YFA0308602). D.S. is supported by the Beijing Nature Science Foundation (JQ19001).
\end{addendum}

\section*{METHODS}

\subsubsection*{Sample growth.}
The single crystals of Cd$_3$As$_2$ were grown by chemical vapor transport method\cite{Chen2015}. Firstly, the polycrystalline of Cd$_3$As$_2$ was synthesized by the mixture of Cd and As powder with the ratio of Cd:As=3:2 and heated at 800 $^{\circ}$C for 10 h. Secondly, the powder of Cd$_3$As$_2$ together with the transport agent I$_2$ (5 mg/ml) was sealed into a quartz tube. Lastly, the ampoule was put into a two-zone furnace with the temperature gradient of 575 $^{\circ}$C and 520 $^{\circ}$C, and kept for 100 h. The the single crystals of Cd$_3$As$_2$ were obtained at the cold end of the tube.

\subsubsection*{ARPES measurements.}
TrARPES measurements were performed in the laboratory at Tsinghua university. A Ti: sapphire oscillator is used as the fundamental laser with tunable wavelength from 710 to 930 nm and repetition rate is reduced to 3.8 MHz by a external pulse picker. Part of the fundamental laser is directly used for pumping and pump fluence is 40 $\mu$J/cm$^2$. The probe laser is generated by fourth harmonics generation using nonlinear crystal BBO and KBBF. The photon energy of probe can be tuned continuously from 5.3 to 7 eV. The beam size is 70 $\mu$m $\times$ 70 $\mu$m for probe and 120 $\mu$m $\times$ 120 $\mu$m for pump. The measurement temperature is 80 K and the vacuum is $<5\times 10^{-11}$ Torr. The overall temporal resolution is 480 fs and energy resolution is 16 meV. The polarizations of pump and probe are $p$-$polarization$.

%
%
%


\end{document}